\documentclass{amsproc}

\newtheorem{theorem}{Theorem}[section]

\theoremstyle{definition}

\theoremstyle{remark}
\newtheorem{remark}[theorem]{Remark}

\numberwithin{equation}{section}
\usepackage{graphicx}

\begin{document}

\title{Statistical Mechanics and Quantum Fields on Fractals}

\author{Eric Akkermans}
\address{Department of Physics, Technion Israel Institute of Technology, Haifa 32000, Israel}
\email{eric@physics.technion.ac.il}
\thanks{The author was supported in part by the Israel Science Foundation Grant No.924/09.}


\date{January 1, 1994 and, in revised form, June 22, 1994.}

\dedicatory{This paper is dedicated to the memory of B. Mandelbrot.}


\begin{abstract}
Fractals define a new and interesting realm for a discussion of basic phenomena  in quantum field theory and statistical mechanics. This interest results from specific  properties of fractals, {\it e.g.}, their dilatation symmetry and the corresponding absence of Fourier mode decomposition. Moreover, the existence of a set of distinct dimensions characterizing the physical properties (spatial or spectral) of fractals make them a useful testing ground for dimensionality dependent physical problems. 
This paper addresses specific problems including the behavior of the heat kernel  and spectral zeta functions on fractals and their importance in the expression of spectral properties in quantum physics. Finally, we apply these results to specific problems such as thermodynamics of quantum radiation by a fractal blackbody.
\end{abstract}

\maketitle

\section{Introduction}

The interest in the behavior of fractals (a word coined by B. Mandelbrot in the 1970's \cite{bookmandel} but without well agreed definition) goes back to the study by mathematicians of strange objects hardly defined by their topology such as the Koch curve or the Sierpinski gasket. These objects are described by continuous but not differentiable functions. At about the same time, probabilists (Levy, Wiener, Doob, Ito, Kolmogorov) and physicists (Smoluchowski, Einstein, Perrin and Langevin to name a few) have been working to give a basis to  the theory of brownian motion, yet another example of fractal object. More recently, physicists have recognized the ubiquitous character of fractals in almost all field of physics, including complex condensed matter (\cite{orbach}-\cite{carlo}), phase transitions \cite{leeuwen,meurice}, turbulence \cite{vaienti}, quantum field theory (\cite{polyakov}-\cite{hill}) and aspects of stochastic processes \cite{sornette,benichou,bernasconi}. An important effort span over more than two decades led to new ideas and concepts to characterize fractals. Notions of self-similarity, iterative maps, fixed points and the identification of distinct fractal dimensions to characterize basic physical properties have been instrumental in the understanding of these objects. 

Since the early 1980's, mathematicians have opened new important directions by being able to define properly brownian motion on some classes of fractals (\cite{barlowperkins}-\cite{derfel}) and Laplacian operators on these structures \cite{kigami,strichartz}. Progress along these two directions have led to a vast literature and it would be a hopeless task to list it exhaustively. Most of these results have been summarized in textbooks (\cite{kigami}-\cite{lapidus}) and reviews \cite{reviewgrabner}. These progresses prove to be  instrumental in physics since they allow to go beyond phenomenological scaling relations towards  a quantitative analysis of fractal structures. 

A useful description in field theory is provided by path integrals and more generally functional integrals \cite{birrell}. Eventually, their evaluation, within one-loop approximation, boils down to the calculation of determinants of operators ({\it e.g.} the Laplacian) and their expression in terms of spectral functions among which the most useful and popular are the heat kernel and corresponding zeta functions (\cite{hawking}-\cite{vassi}). Their evaluation on Euclidean manifolds reveal instrumental in quantum field theory \cite{dunnejphysa}, in statistical mechanics and the theory of phase transitions to name a few \cite{drouffe}. The new tools provided by mathematicians allow to extend these approaches to fractal structures (\cite{eaeuro}-\cite{gdreview}). 

It is the purpose of this paper to give an account of some of these new results. We shall present examples in the realm of  wave and heat propagation on fractals, quantum mechanics, quantum field theory and statistical mechanics.  

Why is it interesting to study physical phenomena on fractal structures ? Fractals define a useful test ground for dimensionality dependent physical phenomena. Indeed, many physical phenomena reveal being critically dependent upon  space dimensionality.  Relevant examples include Anderson localization, Bose-Einstein condensation, onset of superfluidity (Mermin-Wagner-Coleman theorem). On Euclidean manifolds, there is a single space dimensionality so that it is usually not possible to identify the meaning of a dependence upon dimensionality. On a fractal, as we shall see, there exist distinct dimensions which account for geometric, spectral or stochastic informations. A purpose of this paper is to discuss problems where this dependence plays a role.  Fractals also provide a new playground for well designed new experimental setups \cite{dalnegro}. An example is provided by spontaneous emission from atoms embedded in structures whose quantum vacuum has  a fractal  structure \cite{eagurevich}. 
 
\section{Discrete scaling symmetry - Self similarity - Definitions}

\begin{figure}[htb]
\includegraphics[scale=0.65]{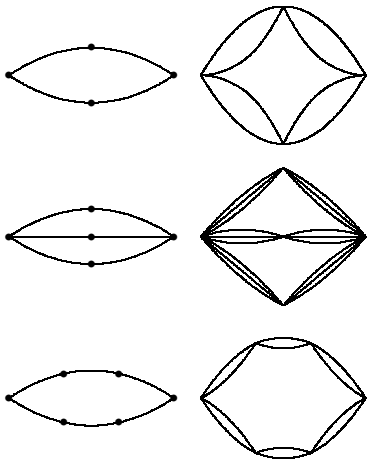}
\caption{First 2 iterations of the diamond fractals $D_{4,2}$, $D_{6,2}$ and $D_{6,3}$.  Their respective branching factors (defined in the text) are $B = 1,2,1$. }
\label{figfract}
\end{figure}

As opposed to Euclidean spaces characterized by translation invariance, self-similar (fractal) structures possess a dilatation symmetry of their physical properties, each characterized by a specific fractal dimension. To illustrate our results, we shall consider throughout this paper simple examples of fractals such as the Siepinski gasket, families of diamond fractals represented on Fig.\ref{figfract} or Cantor sets, but the reader should keep in mind that our results apply to a broader class of self-similar deterministic fractals. Much less is known about other related but more complex systems including random fractals ({\it e.g.} critical percolation clusters), multifractals, T-graphs, trees or more generally objects which do not exhibit an exact decimation symmetry like in deterministic fractals considered here \cite{gefen}.

A fractal is an iterative structure. Let us give a simple example of dimensional characterization. To that purpose, consider a triadic Cantor set obtained iteratively by removing from a given initial interval of length $L_0$ the middle third part.  Define a uniform mass density on this interval so that the initial mass is $M_0$ for a length $L_0$. After the first iteration, the mass is $2 M_0$ while the length is $3 L_0$. After the $n$-th iteration, the mass is $M_n = 2^n M_0$ while the length becomes $L_n = 3^n L_0$, so that the fractal dimension describing the mass density, the geometric Hausdorff dimension $d_h$, is given by,
\begin{equation}
d_h = \mbox{lim}_{n \rightarrow \infty} {\ln M_n \over \ln L_n } = {\ln 2 \over \ln 3}
\label{def1}
\end{equation}

An alternative way to obtain this result, is to start from the scaling relation between masses $M(L)$ at different lengths $L$, namely $M(L) = M_0 (L) + {1\over 2} M(3L)$ for the triadic Cantor set, where $M_0 (L)$ is the initial mass. More generally, we are interested in the solution of the equation 
\begin{equation} 
f(x) = g (x) + { 1 \over b} f(a x)
\label{scaling}
\end{equation}
with determined scaling parameters $a$ and $b$ and a given initial function $g (x)$. This form defines a {\it discrete scaling symmetry} as opposed to continuous scaling which extends (\ref{scaling}) to any rescaling $a$ of the variable $x$. 
For the specific case $g = 0$, a solution of (\ref{scaling}) can be sought under the form $f(x) = x^\alpha F(x)$ for generally non constant $F(x)$. Inserting into (\ref{scaling}), and choosing $\alpha = \ln b / \ln a$, we obtain that $F$ must fulfill $F(ax) = F(x)$. Redefining $F(x) \equiv G ( \ln x / \ln a )$ leads to $G ( \ln x / \ln a  +1 ) = G ( \ln x / \ln a )$, namely  the general solution of (\ref{scaling}) is of the form,
\begin{equation}
f( x) = x^{\ln b / \ln a } \, G \left( {\ln x \over \ln a } \right)
\label{scalingsol}
\end{equation}
where $G$ is a periodic function of its argument of period unity. The scaling form (\ref{scaling}) can be iterated so that it rewrites
\begin{equation}
f(x) = \sum_{n=0}^\infty \, b^{-n} \, g(a^n \, x) \, .
\label{iteration}
\end{equation}
This form makes explicit an important feature of  discrete scaling symmetry namely the scaling function $f(x)$ can be written as a series with exponentially growing coefficients $b^{-n}$ and $a^n$ instead of polynomial growth. This is an essential and ubiquitous feature of self-similarity  that we shall encounter all along this paper. 

\subsection{Mellin transform}
To study functions involving dilatation scaling symmetry, it is desirable to find an appropriate transform equivalent to Laplace or Fourier transforms for translation symmetry. The Mellin transform plays this role \cite{frenchreviewmellin}. For a function $f(t)$ defined on the positive real axis $0 < t < \infty$, its Mellin transform 
\begin{equation}
M_f (s) \equiv \int_0 ^\infty dt \, t^{s-1}  \, f(t) 
\label{mellin18}
\end{equation} 
is defined on the complex plane. The Mellin transform of $f(t)$ is the Laplace transform of the function $g(t) \equiv f \left( e^{-t} \right)$, a property which precisely accounts for the dilatation symmetry in (\ref{scaling}) or the exponential behavior in (\ref{iteration}). An important property of the Mellin transform can be stated as follows. If $f(t)$ is analytic in $0 < t < \infty$ and $f(t) = O( t^{- \alpha} )$ for $t \rightarrow 0$ and  $f(t) = O( t^{- \beta} )$ for $t \rightarrow \infty$, with $\alpha < \beta$, then the Mellin transform $M_f (s)$ is analytic in the strip $\alpha < \mbox{s} < \beta$, and $f(t) = (1 / 2i \pi ) \int_{\sigma - i \infty}^{\sigma + i \infty} M_f (s) \, t^{-s} \, ds$ where $\alpha < \sigma < \beta$. Another suitable expression of the Mellin transform is provided by the zeta function of the function $f(x)$ defined by 
\begin{equation}
\zeta_f (s) \equiv { 1 \over \Gamma (s)} \int_0^\infty dx \, x^{s-1} \, f(x)
\label{zeta1}
\end{equation}
which takes into account suitable properties of the Euler $\Gamma$ function as we shall see later. 
The inverse transform is thus
\begin{equation}
f(x) =  { 1 \over 2 i \pi} \int_{\sigma - i \infty}^{ \sigma + i \infty} ds \, x^{-s} \Gamma (s) \,  \zeta_f (s) 
\label{inversemellin}
\end{equation}
A direct calculation of the zeta function of a function $f(x)$ with a discrete scaling symmetry (\ref{scaling}) is 
\begin{equation}
\zeta_f (s) = { b \, a^s \, \zeta_g (s) \over  b\, a^s -1}
\label{zeta2}
\end{equation}
The behavior of $f(x)$ is driven by the poles of $\zeta_f$. Disregarding at this stage the pole structure of $\zeta_g$, we focus on the poles in (\ref{zeta2}) which result  from the scaling symmetry. There are the solutions, $s_n$, of $b \, a^s = 1$ namely, 
\begin{equation}
s_n = { \ln \left( 1 / b \right) \over \ln a} + {2 i \pi \, n \over \ln a}
\label{complexpoles}
\end{equation}
for integer $n$. The origin of these poles is a direct consequence of the exponential behavior of the coefficients in (\ref{iteration}). These complex poles have been identified with complex valued dimensions of self-similar fractal systems \cite{lapidus}. By an inverse transform, we have 
\begin{equation}
f(x) = x^{ - { \ln \left( 1/b \right) \over \ln a}} \sum_{n=-\infty}^{+ \infty} \Gamma(s_n) \, \zeta_g (s_n) \, e^{ -2 i \pi n \, \left( {\ln x \over \ln a} \right)} \equiv x^{ - { \ln \left( 1/b \right) \over \ln a}} G \left( {2 \pi \ln x \over \ln a} \right) 
\label{logperiodic1}
\end{equation}
namely, the general solution of (\ref{scaling}) is the product of a power law behavior characterized by a fractal dimension $  \ln \left( 1/b \right) / \ln a$ times a periodic function $G(x+1) = G(x)$ of its argument $2 \pi \ln x / \ln a$. This log-periodic behavior is tightly related to the existence of a discrete scaling symmetry as expressed by (\ref{scaling}) and it constitutes its fingerprint together with the power law prefactor determining the fractal dimension. 

Log-periodic functions have a long history \cite{valiron} including the well known Weierstrass function determined by the series,
\begin{equation}
W (t) = \sum_{n=0}^\infty a^n \, \cos \left( b^n \, t \right) \, . 
\label{w}
\end{equation}
 They appear in a broad range of problems such as renormalization group and the theory of phase transitions \cite{leeuwen,meurice}, Markov processes in complex media \cite{bernasconi}, turbulence \cite{vaienti} and fractals \cite{bessis,sornette}. Let us also mention that the scaling relation (\ref{scaling}) with definite parameters $(a,b)$ is a particular example of more general iteration processes described by the Poincar\'e equation $f( az) = {\mathcal P} \, \left( f(z) \right)$ where ${\mathcal P} (x)$ is a polynomial \cite{valiron}. This is actively studied in the mathematical literature \cite{reviewgrabner}.

\begin{remark}
The periodic function $G(x)$ becomes constant when all residues in the inverse Mellin transform vanish except for $n=0$. This is the case for instance for the mass density $M(L)$ of the interval of length $L$ or generally of any Euclidean manifold of dimension $d$, where $M(L) \propto L^d$ is expected without log-periodic behavior. This can be checked either by a direct calculation or by noticing that if the scaling symmetry $M(a L ) = b \,  M(L)$ takes place for any value of $a$ and not only for a fixed one, then, averaging over $a$ washes out the oscillations leaving only the expected power law behavior. 

\end{remark}

\subsection{A variational derivation}

It is of interest to present another derivation of the form (\ref{logperiodic1}) not based on the inverse Mellin transform. This provides another point of view which is useful in cases where the pole structure cannot be easily retrieved. To that purpose, we start from the iterated form (\ref{iteration}) of $f(x)$ and to estimate it, we use a saddle point approximation to find the value of $n$ which dominates $f(x)$. By differentiating with respect to $n$, we obtain $- (\ln b) g + (\ln a) a^n \, x \, g' =0$. Defining $u = a^n \, x$, we have $ - (\ln b) g(u) + (\ln a) u \, x \, g' (u) =0$. This equation in the variable $u$ admits a solution $N$ such that $u = a^N \, x$ and then,
\begin{equation}
N = { \ln u - \ln x \over \ln a } \, .
\label{u}
\end{equation}
$N$ is not necessarily an integer, namely $N = n_0 + t$, where $n_0$ is an integer and $0 \leq t \leq 1$. Then, we write $f(x)$ in (\ref{iteration}) under the form,
\begin{equation}
f(x) = \sum_{m= - n_0 +1}^\infty b^{- (n_0 + m)} \, g \left( a^{n_0 + m} \, x \right) 
= \sum_{m= - n_0 +1}^\infty b^{- (N + m - t)} \, g \left( a^{N + m -t} \, x \right) \, .
\label{u2}
\end{equation}
Using (\ref{u}) leads to $b^{-N} = \left( u / x \right)^{\ln (1/b) / \ln a}$, and since, $u$ and $a$ are fixed quantities, then $n_0 \rightarrow - \infty $ for $x \rightarrow 0$, so that
\begin{equation}
f(x) = \left( { u \over x } \right)^{ {\ln (1/b) \over \ln a}} \sum_{m= - \infty}^\infty b^{- (m - t)} \, g\left( a^{m-t} \, u \right) 
\end{equation}
where the series is now a periodic function of $t$ of period unity, namely using (\ref{u}), a periodic function of $\ln x / \ln a$ of period unity. We thus recover (\ref{logperiodic1}).

\section{Heat kernel and spectral  functions - Generalities}

The study of spectral functions on manifolds has a long and very successful history which traces back to the early XXth century with considerations put forward by Lorentz on the blackbody radiation \cite{reviewweyl}. Given an Euclidean manifold, it is possible to retrieve some of its geometric characteristics such as its volume, surface, curvature, Euler-Poincare characteristics by studying the spectral properties of the Laplace operator defined on the manifold \cite{reviewweyl,baltes76,molchanov}. In other words, the Laplacian or more precisely its eigenvalue spectrum  can be viewed as  a "ruler" which allows to span the manifold. Since stationary wave and heat equations are both governed by the Laplacian, looking at the heat flow or at (scalar) electromagnetic  wave propagation are equivalent ways for extracting geometric characteristics of Euclidean manifolds. This has led to seminal works in mathematics starting with H. Weyl \cite{weyl} and the expansion which bears his name culminating in the celebrated question of M. Kac, "Can you hear the shape of a drum ?" \cite{kac} and the negative answer provided by examples of isospectral domains of different shapes \cite{rosenberg}. 

The direct relation between spectral functions of Laplace operators and path integral in quantum field theory \cite{dunnejphysa} or equivalent functional forms of the partition function in statistical mechanics are the underlying reasons for the enduring success of  methods based on spectral functions of Laplace or corresponding Dirac operators in gauge theories \cite{atiyah}. 

\medskip

Let us first illustrate these ideas using simple examples (see Chapter 5 in \cite{ambook}). Consider first the diffusion of heat $\phi (x,t)$ along the infinite, unbounded real line $(d=1)$. The corresponding heat equation is 
\begin{equation}
{\partial \phi \over \partial t} = D \Delta \phi 
\label{heat1}
\end{equation}
where the diffusion coefficient $D$ sets units of length and time. The Green's solution of this equation is 
\begin{equation}
P_{d=1} (x,y,t) = {1 \over \left( 4 \pi D t \right)^{1/2} } e^{- { (x-y)^2 \over 4 Dt}} \, .
\label{greenheat}
\end{equation}
In a probabilistic interpretation, $P_{d=1} (x,y,t)$ represents the probability density for a particle to diffuse from an initial position $x$ to a final position $y$ in a time $t$. Its generalization $P_d (x,y,t)$ to the $d$-dimensional free space is obtained from (\ref{greenheat}) by replacing the exponent $1/2$ in the denominator by $d/2$. A way to characterize the space geometry which sustains the heat flow, is to consider the heat kernel defined by 
\begin{equation}
Z_d (t) = \int_{V} d^d x \, P_d (x,x,t) = {V \over  \left( 4 \pi D t \right)^{d/2} }
\label{heat17}
\end{equation}
where the integral is over a volume $V$ defined qualitatively without yet specifying boundary conditions. The heat kernel thus defined is an integral over all closed diffusing trajectories (starting and ending at a point $x$) within the volume $V$.  The Green's function $P_d$ is obtained for $t >0$, from the normalized eigenfunctions $\phi_n (x)$ and non negative eigenvalues $\lambda_n$ (with degeneracy $g_n$) of the heat equation (\ref{heat1}) in $d$ dimensions, as 
\begin{equation}
P_d (x,y, t) = \sum_n g_n \,  \phi_n (x) \, \phi^* _n (y) \, e^{- \lambda_n \, t}
\label{greenheat2}
\end{equation}
so that 
\begin{equation}
Z_d (t) =  \sum_n g_n \, e^{- \lambda_n \, t} = \mbox{Tr} \left( e^{D\, t \, \Delta} \right) \, . 
\label{heat2}
\end{equation}
This relation between the heat kernel $Z_d (t)$ and the Laplace operator $- \Delta$, expresses the distribution of closed diffusive trajectories within a manifold in terms of the spectrum of the Laplacian. This relation for the heat kernel is instrumental in  calculating Euclidean path integrals \cite{dunnejphysa}, partition functions \cite{drouffe} and other spectral and transport quantities \cite{ambook}. 

\subsection{The Weyl expansion}

Expression (\ref{heat2}) is also very convenient to evaluate $Z_d (t)$ for manifolds with boundaries. Consider the simple case of diffusion on an  interval of length $L$. The corresponding eigenvalue spectrum of the Laplacian (we set $D=1$ for convenience) is given by $\lambda_n = \left( n \pi / L \right)^2$ where $n$ is an non zero integer for Dirichlet $(\mathcal D)$ boundary conditions, $\phi (x=0,t) = \phi (L,t) = 0$, whereas it includes the zero mode $n=0$ for Neumann $(\mathcal N)$ boundary conditions $\partial \phi (x,t) |_{x=0} = \partial \phi (x,t) |_{x=L} = 0$. The corresponding heat kernels $Z_{\mathcal N,\mathcal D} (t)$ are thus related by 
\begin{equation}
Z_{\mathcal N} (t) = \sum_{n=0}^\infty e^{- \left(n \pi / L \right)^2 \, t } = 1 + Z_{\mathcal D} (t) 
\label{zd}
\end{equation}
The use of the Poisson formula allows to write the small time asymptotic expansion 
\begin{equation}
Z_{\mathcal N,\mathcal D} (t) = {L \over \sqrt{4 \pi  \, t}}  \mp {1 \over 2} + \cdots
\label{heat1d}
\end{equation}
This is the simplest example of a Weyl expansion. For a two-dimensional domain of arbitrary shape with surface $S$ and boundary length $L$, we have the Weyl expansion corresponding to Dirichlet boundary conditions \cite{waechter},
\begin{equation}
Z_2 (t) = {S \over 4 \pi \,  t} - { L/4 \over \sqrt{4 \pi \, t} } + {1 \over 6} + \cdots 
\label{weyl2d}
\end{equation}
where the constant term, $1/6$, results from the integral of the local curvature of the boundary. More generally for a $d$-dimensional Euclidean manifold of hypervolume $V$, hypersurface $S$, {\it etc.}, the Weyl asymptotic expansion (restoring the diffusion coefficient $D$), involves powers of $1 / \left( 4 \pi \, Dt \right)^{(d-i) /2} $: 
\begin{equation}
Z_d (t) \simeq {V \over (4 \pi D \, t)^{d/2}} - \alpha_d {S \over (4 \pi D \, t)^{(d-1)/2}} + \cdots
\label{weyld}
\end{equation}
where $\alpha_d$ is a constant which depends on boundary conditions \cite{vassi,reviewweyl,baltes76}. The Weyl asymptotic formula provides a small time expansion for $Z_d (t)$. Physically, it describes the behavior of a diffusive particle initially released at some point in the manifold. At small time, it experiences a free space diffusion insensitive to the boundaries (volume term). At later times, the particle starts feeling the boundary (surface term), then its shape (local curvature term), {\it etc.}.  

To conclude this part, we discuss a point concerning the dispersion relation ({\it i.e.}, the relation between time and length units) which will prove relevant when considering diffusion on fractals. 
 $Z_d$ is, as defined in (\ref{heat2}), a dimensionless quantity. On the other hand, the Laplacian has dimensions of the inverse of a length squared which allows to retrieve from the heat kernel  geometric information about the manifold. We thus need to insert a diffusion coefficient $D$ in order for  $V^{2/d} / D$ in (\ref{weyld}) to have units of time. The diffusion coefficient $D$ expresses the underlying physics of the diffusion flow, and it is related to the relevant sources of diffusion by an appropriate phenomenological relation ({\it e.g.} temperature $T$ and viscosity $\sigma$ where $D = R \, T / 6 \pi \, \eta \, a \, \mathcal N = k_B T \sigma$) generally known as the Einstein relation \cite{bookmecastat}. For instance, in the specific case of a covariant diffusion equation as studied in quantum mesoscopic physics \cite{ambook}, or in Euclidean time formulation of the Schr\"odinger equation of a particle of mass $m$, we have $D = \hbar / 2m$.

\subsection{Spectral determinant - Density of states and spectral zeta function}

There is an important and useful relation between the heat kernel, its geometrical content and the density of states $\rho (\lambda)$ of the Laplacian defined on the corresponding manifold. 
To find it, we define the spectral determinant associated to the eigenvalue spectrum $\lambda_n$ of $- \Delta$, 
\begin{equation}
S(\gamma) = \mbox{det} \left( - D \Delta + \gamma \right) = \prod_n \left( \lambda_n + \gamma \right) 
\label{det}
\end{equation}
where $\gamma$ is a real number. From the relation (\ref{heat2}), it follows that
\begin{equation}
\int_0^\infty Z(t) e^{-\gamma t} dt = \sum_n {g_n \over \gamma +
E_n} = {\partial \over \partial \gamma} \mbox{ln} S(\gamma) \ .
\label{detspec2} 
\end{equation}
The density of states $\rho ( \lambda ) = \sum_n \delta ( \lambda - \lambda_n )$ is thus directly related to the heat kernel and the spectral determinant through (see for instance Chapter 5 in \cite{ambook})
\begin{equation}
\rho (\lambda) = - {1 \over \pi} \mbox{lim}_{\eta \rightarrow
0^{+}} \mbox{Im} {d \over d \gamma } \mbox{ln} S(\gamma)
\label{dosdet} \end{equation} where $\gamma$ is now complex valued, 
$\gamma = - \lambda + i \eta$. This last relation proves useful but uneasy to implement since the spectral determinant $S(\gamma)$ is defined by the product of
eigenvalues $\lambda_n$. This product is infinite,
so that its definition is formal. To give an interpretation to
$S(\gamma)$, we resort to the spectral $\zeta_\Delta$
function, associated to the
Laplacian and defined by 
\begin{equation} 
\zeta_\Delta (s) =
\sum_{n} {1 \over \lambda_n^s} \, .
 \label{zeta20} \end{equation}
 This function is well-defined for
all values of $s$ for which the series converges. Using the
identity
\begin{equation} {1 \over \lambda^{s}} = {1 \over \Gamma (s)} \int_{0}^{\infty}
dt \ t^{s-1} e^{- t \lambda} \ \ , \end{equation}
we write
\begin{equation} \zeta_\Delta (s) = {1 \over \Gamma (s)} \int_{0}^{\infty} dt \
t^{s-1} \mbox{Tr} \left( e^{D\, \Delta \, t} \right) \ \ . \label{mellin1} \end{equation}
$\zeta_\Delta (s)$ thus defined, is the Mellin
transform (\ref{mellin18}) of the heat kernel. It is convergent, and its analytic
continuation in the complex plane defines a meromorphic function
in $s$ which is analytic at $s=0$. We use this analyticity and the
identity~:
\begin{equation} {d \over ds} \lambda_n^{-s}|_{s=0} = - \mbox{ln} \lambda_n
\label{zeta12}\end{equation}
to express the spectral determinant as \begin{equation} \displaystyle \mbox{ln}
S(\gamma = 0) = - {d \over ds} \zeta_\Delta (s) |_{s=0}
\label{mellin2} \end{equation} which is well defined.

As an example, we consider the Laplacian on an interval of length $L$ with Dirichlet boundary conditions (and set $D=1$). Inserting the expression $\lambda_n = \left( n \pi / L \right)^2$ of the eigenvalues into (\ref{zeta20}), we obtain
\begin{equation}
\zeta_\Delta (s) = \sum_{n=1}^\infty \left( { L^2 \over \pi^2 n^2 } \right)^s = \left( {L \over \pi } \right)^{2s} \zeta_R (2s) \, .
\label{zetadelta2}
\end{equation}
The Riemann zeta function $\zeta_R (2s)$ has a simple pole at $2s =1$ and more generally a simple pole at $2s =d$ for a $d$-dimensional Euclidean manifold. Thus, the inverse Mellin transform (\ref{inversemellin}) provides directly $Z_{\mathcal D} (t) = \left( L / 2 \pi \right) \Gamma (1/2) t^{- 1/2} + \cdots$ namely (\ref{heat1d}). Using (\ref{dosdet}), we deduce the main contribution to the corresponding density of states
\begin{equation}
\rho_{d=1} (\lambda) \simeq {L \over 2 \pi \, \sqrt{ \lambda}}
\label{dos1d}
\end{equation}
and from (\ref{weyld}), we obtain the generalization of these results to $d$-dimensional Euclidean manifolds, 
\begin{eqnarray}
\rho_3 (\lambda) &=& {V \over 4 \pi^2}  \sqrt{\lambda}  - {S \over 16 \pi }  + \cdots \nonumber \\
\rho_2 (\lambda) &=& {S \over 4 \pi}  - {L \over 8 \pi } {1 \over \sqrt{\lambda}} + \cdots
\label{dosd}
\end{eqnarray}
To conclude this part, we note that the short time Weyl expansion of the heat kernel is related to the pole structure of the zeta function $\zeta_\Delta$ associated to the Laplacian.  

\subsection{Counting function - Spectral zeta function - Wavelet transform}

We consider  the counting function $N(\lambda)$ defined by 
\begin{equation}
N(\lambda) = \sum_{\lambda_n < \lambda} 1 \, .
\label{counting}
\end{equation}
The counting function $N(\lambda)$ and the density of states $\rho (\lambda) = {d N(\lambda) \over d \lambda} = \sum_n \delta (\lambda - \lambda_n )$ are the sum of step $\theta$ and Dirac $\delta$ functions characterizing the distribution of eigenvalues $\lambda_n$ of the Laplacian. The density of states $\rho (\lambda)$ is related to the spectral determinant (\ref{dosdet}), to the heat kernel and to the spectral zeta function. Unlike the counting function, the density of states is often ill-defined since $N(\lambda)$ is not differentiable ({\it e.g.} for a fractal spectrum).  

Consider the number $C(\lambda)$ of eigenvalues of the Laplace operator $- \Delta$ within the interval $\left[ 0, \lambda \right]$ included into the support $J$ of the whole spectrum. It is given in terms of the counting function as 
\begin{equation}
C(\lambda ) = N(\lambda) - N(0) = \int_J dN (\epsilon) \, \theta \left( \lambda - \epsilon \right)
\label{counting1}
\end{equation}
where the integral extends over the whole spectrum $J$ with the spectral measure $dN (\epsilon)$. By definition (\ref{heat2}), the heat kernel $Z(t)$ can be expressed as (with $D=1$)
\begin{equation}
Z(t) = \mbox{Tr} \left( e^{t \, \Delta} \right) = \int_J dN(\lambda) \, e^{-t \, \lambda} \, . 
\label{heatcounting}
\end{equation}
The spectral zeta function $\zeta_\Delta (s)$ defined in (\ref{mellin1}) thus becomes,
\begin{eqnarray}
\zeta_\Delta (s) &=& {1 \over \Gamma (s) } \, \int_0^\infty dt \, t^{s-1} \int_J dN(\lambda) \, e^{-t \, \lambda} \nonumber \\
&=& \int_J dN (\lambda) \, \lambda^{-s}
\label{zetan}
\end{eqnarray}
as a result of the identity $\Gamma (s) = \int_0^\infty dx \, x^{s-1} \, e^{-x}$. It is thus possible to relate the Mellin transform $M_C (s)$ of $C(\lambda)$ to $\zeta_\Delta (s)$, namely,
\begin{eqnarray}
M_C (s) &=& \int_J \, dC(\lambda) \, \lambda^{-s} = \int_J \lambda^{-s} \, d\left( \int_J dN (\epsilon) \, \theta \left( \lambda - \epsilon \right) \right) \nonumber \\
&=& \int_J \lambda^{-s} \, \int_J dN(\epsilon ) \, \delta (\lambda - \epsilon ) \nonumber \\
&=& \zeta_\Delta (s)
\label{zetac}
\end{eqnarray}
so that the spectral function $ \zeta_\Delta (s)$ is the Mellin transform of the counting function. This relation is a particular example of a more general property of the Mellin transform which can be viewed as a convolution theorem. To state it, consider  an integrable function $g(x)$, suitably normalized to unity, $\int_0^\infty dt \, g(t) =1$, and whose Mellin transform $M_g (s)$ is defined. The quantity 
\begin{equation}
W_g \left( \lambda_b , t \right) \equiv \int_J dN(\lambda) \, g \left(t \, |\lambda - \lambda_b | \right) \, ,
\label{wavelet}
\end{equation}
defined for $t > 0$, is called the wavelet transform of the counting function $N(\lambda)$. Intuitively, the wavelet transform can be viewed as a mathematical microscope which probes the counting function at a point $\lambda_b$ with an optics specified by the choice of the specific wavelet $g(x)$. An important property of the wavelet transform is that it preserves a discrete scaling symmetry (\ref{scaling}) of the probed function. Performing a Mellin transform of $W_g \left( \lambda_b , t \right)$ {\it w.r.t.} the variable $t$ gives
\begin{equation}
M \left[ W_g \left( \lambda_b , t \right) \right] = \zeta_\Delta \left( s, - \lambda_b \right) \, M_g (s)
\label{mellinwavelet}
\end{equation}
where $\zeta_\Delta \left( s, - \lambda_b \right) \equiv \int_J dN(\lambda) \, |\lambda - \lambda_b |^{-s}$ is a shifted zeta function. From the last relation, it is immediate to obtain the relation between different wavelet transforms respectively specified by the "optics" $f(t)$ and $g(t)$ of Mellin transforms $M_f $ and $M_g$, namely
\begin{equation}
{M \left[ W_g \left( \lambda_b , t \right) \right] \over M_g (s) } = {M \left[ W_f \left( \lambda_b , t \right) \right] \over M_f (s) } = \zeta_\Delta \left( s, - \lambda_b \right) \, .
\label{convolution}
\end{equation}
An interesting application of this convolution rule is obtained by taking $\lambda_b =0$ and the wavelet $g(t) = { 2\sin t \over \pi t}$ such that $M_g (s) = (2 / \pi) \, \Gamma(s-1) \, \sin \pi (s-1)/2$ for $0 < \mbox{Re} (s) < 2$. We obtain from (\ref{wavelet}), 
\begin{equation}
M \left[ {2 \over \pi} \int_J dN(\lambda) \, {sin \lambda \, t \over \lambda \, t}  \right] = {2 \over \pi}  \Gamma(s-1) \, \sin {\pi \over 2} (s-1) \,  \zeta_\Delta (s)
\end{equation}
so that from the inverse Mellin transform,
\begin{equation}
{2 \over \pi} \int_J dN(\lambda) \, {sin \lambda \, t \over \lambda \, t} 
 = {1 \over 2 i \pi} \int_{\sigma - i \infty}^{\sigma + i \infty} ds \, t^{-s} \, \mbox{sinc} \left[ {\pi (s-1) \over 2} \right] \, \Gamma(s) \, \zeta_\Delta (s) \, .
\label{convolution2}
\end{equation}
Therefore different probes of the spectrum of the Laplacian can be related one to another and expressed in terms of the spectral zeta function. This is a powerful result with interesting physical consequences since distinct  physical phenomena  usually involve their own probing function ({\it e.g.} the  Fermi golden rule describing the response of the spectrum to an external perturbation (for instance spontaneous emission), involves the sinc probe $g(t)$) \cite{eagurevich} and all are related to the heat  kernel characterized by an exponential probe.


\section{ Laplacian on fractals - Heat kernel and spectral zeta function}

This section is devoted  to  studying the heat kernel on fractals, namely on systems whose geometry is characterized by a discrete scaling symmetry as defined in (\ref{scaling}). In other words, we would like to extend the  previous analysis and the relation between geometric and spectral properties to fractals. This  is a vast subject and we do not intend to be exhaustive but rather to study specific but generic enough 
 examples in order to highlight some of the more salient results and open questions. 
 
 We have seen that  geometric characteristics of Euclidean manifolds can be retrieved from the spectrum of the corresponding Laplace operator $ - \Delta$. The relevant spectral tools are the heat kernel (\ref{heat2}) and the  spectral zeta function $\zeta_\Delta (s)$ defined by  (\ref{mellin1}).  How do they generalize to fractals ? Geometric information about a fractal is characterized by its Hausdorff dimension $d_h$. On the other hand, the very notion of volume or surface of a fractal is rather ill-defined. Then, is it possible to define a heat kernel and a Weyl expansion for fractals and if it is so, what kind of information does it provide. Those questions have been and still are in the focus of intensive works from the mathematical community where an abundant number of important results has been proved (\cite{barlowperkins}-\cite{reviewgrabner}). It is not our purpose to review them, but rather to show their relevance and usefulness in   physics. An important step  has been to prove, using either a probabilistic or an analytic approach, that a Laplace operator can be properly defined on fractal structures \cite{kigami} as well as corresponding heat kernel and spectral zeta-function \cite{teplyaev}.
  
 As we have seen in Section 2, the characteristic feature of a discrete scaling symmetry is the existence of a tower of complex poles in the zeta function associated to the relevant quantity. These complex poles are a direct expression of the scaling form (\ref{scaling}) and of the exponential behavior of the coefficients in the iteration series (\ref{iteration}). Are there  similar properties of the spectral zeta function of the Laplacian on a fractal. The answer to this question is positive. There exist two characteristic parameters $(a,b) \equiv  (l^{d_w} , l^{d_h})$, such that the heat kernel of the Laplacian on a self-similar fractal obeys the scaling relation (\ref{scaling}). These scaling parameters involve the geometric Hausdorff dimension $d_h$, the {\it walk dimension} $d_w$ that we shall define and discuss later on and an elementary step length $l$ describing size scaling.

\begin{figure}[hbt]
\begin{center}
\begin{tabular}{||l||r|r|r|r||}
\hline 
& $d_h$ \, \  & $d_w$ \, \ & $d_s = 2 d_h / d_w$ & $l=L_n ^{1/n}$ \\ 
\hline  
 \, \ \ $D_{4,2}$ & $2$ \, \ \   & $2$ \, \ \
& $2$ \, \ \ & $2$ \, \ \  \\
\hline 
 \, \ \  $D_{6,2}$ & ${\ln 6 / \ln 2} $ & $2$ \, \ \ & ${\ln 6 / \ln 2}$ \, & $2$ \, \ \ \\
\hline
 \, \ \  $D_{6,3}$ & ${\ln 6 /  \ln 3} $ & $2$ \, \ \ & ${\ln 6 / \ln 3} $ \, & $3$ \, \ \ 
\\
\hline 
  $\mbox{Sierpinski}$ & ${\ln 3 / \ln 2}$ & ${\ln 5 / \ln 2}$ &
 $2 \, {\ln 3 / \ln 5}$ \, &
$2$ \, \ \ \\
\hline 
\end{tabular}
\caption{Fractal dimensions and size scaling factor for diamond fractals and for the Sierpinski gasket. For  $D_{6,2}$, the spectral dimension is $d_s\approx 2.58$, and for $D_{6,3}$, $d_s\approx 1.63$.}
\end{center}
\label{table}
\end{figure}

 To proceed further, we consider the specific case of a diamond fractal $(\Diamond)$ (see Figs.\ref{figfract} and \ref{table}).  It has been shown \cite{eaeuro} that the corresponding heat kernel $Z_{\Diamond} (t) $ has a closed expression of the form (\ref{iteration}) where the initial function to be iterated is the heat kernel $Z_{\mathcal D} (t)$ of the interval of unit length with Dirichlet $( \mathcal D )$ boundary conditions, given by (\ref{zd}), namely,
$Z_{\mathcal D} (t) = \sum_{k=1}^\infty e^{- k^2 \pi^2 t}$ (we set $D=1$ for convenience), so that,
\begin{equation}
Z_{\Diamond} = Z_{\mathcal D} (t) + B \sum_{n=0}^\infty L_n ^{d_h} Z_{\mathcal D} \left( L_n ^{d_w} \, t \right)
\label{zdiamond}
\end{equation}
where we have defined the total length $L_n \equiv l^n$ of the diamond fractal at step $n$ of the iteration. The coefficient $B \equiv \left( l^{d_h -1} \, -1 \right)$ is the branching factor of the fractal (see Fig.\ref{figfract}) and the integer $l^{d_h}$ is the number of links into which a given link is divided. As in (\ref{iteration}), the series for $Z_\Diamond (t)$ involves coefficients which behave exponentially with the iteration $n$. We therefore expect an expression of the form (\ref{logperiodic1}) with log-periodic oscillations with time. To show it, we calculate the corresponding spectral zeta function, 
\begin{equation}
\zeta_\Diamond = { 1 \over \Gamma (s) } \int_0 ^\infty dt \, t^{s-1} \, Z_\Diamond (t) \, .
\label{zetadiamond}
\end{equation}
An elementary calculation \cite{eaeuro} leads to 
\begin{eqnarray}
\zeta_\Diamond (s) &=& {\zeta_R(2s) \over \pi^{2s}} \left( 1 + B \sum_{n=0}^\infty L_n^{{d_h}  - d_w s} \right)
 \nonumber \\
&=& {\zeta_R(2s) \over \pi^{2s}} l^{{d_h} -1} \left( {1 - l^{1 - d_w s} \over 1 - l^{{d_h} - d_w s} } \right)\quad ,
\label{zeta3}
\end{eqnarray}
where $\zeta_R(2s)$ is the Riemann zeta function. Note that a similar structure exists for the Sierpinski gasket \cite{teplyaev,strichartz}, with the Riemann zeta function factor replaced by another zeta function.
$\zeta_\Diamond (s)$ has complex poles given by
\begin{equation}
s_m = \frac{{d_h}}{d_w} + {2 i \pi m \over d_w \ln l} \equiv \frac{d_s}{2}+ {2 i \pi m \over d_w \ln l}\quad ,
\label{poles}
\end{equation}
where $m$ is an integer. These complex poles have been identified with complex dimensions for fractals \cite{lapidus}. The fractal dimension $d_s \equiv 2 d_h / d_w$ is called {\it spectral dimension}. It has been obtained in earlier works in the physics community \cite{rammal,rammaljphys} and recognized as the relevant fractal dimension (unlike $d_h$) underlying spectral properties of self-similar fractals.

\begin{figure}[ht]
\includegraphics[scale=0.35]{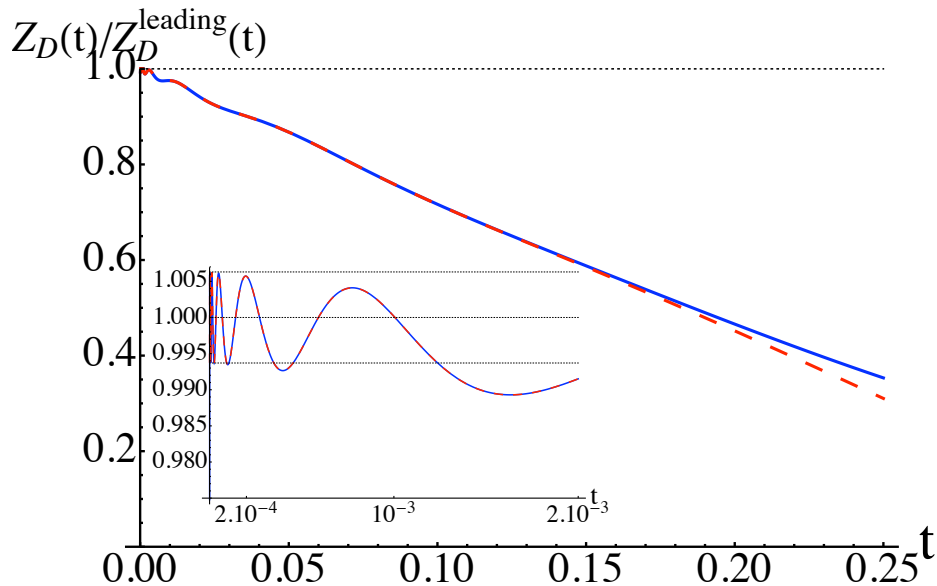}
\caption{Heat kernel $Z_D (t)$ at small time, normalized by the leading non-oscillating term, for the fractal diamond $D_{4,2}$.  The solid [blue] curve  is exact; the dashed [red] curve is the approximate expression  (\ref{zdiam2}). At very small $t$, these curves are indistinguishable, as shown in the inset plot. The relative amplitude of the oscillations  remains constant as $t\to 0$ \cite{eaeuro}.}
\label{figosc}
\end{figure}

From the inverse Mellin transform (\ref{inversemellin}), we obtain 
\begin{eqnarray}
Z_\Diamond (t) &=& {V_s \over t^{d_s /2}} \sum_{m= \, - \infty}^\infty \Gamma (s_m) \zeta_R(2s_m)/ \pi^{2s_m} e^{2 i \pi \, m \, \ln t / (d_w \, \ln l )} \nonumber \\
&=&  {V_s \over t^{d_s /2}} G_\Diamond \left( { 2 \pi \, \ln t \over d_w \, \ln l } \right) 
\label{zdiam2}
\end{eqnarray}
where $G_\Diamond$ is a periodic function of its argument of period unity. These log-periodic oscillations are represented on Fig.\ref{figosc} and we note that the higher order (with $m$) complex poles give much smaller contributions, a result related to the steep decrease of the Euler $\Gamma$ function. A similar behavior has been found numerically for the Sierpinksi gasket \cite{strichartz}. 

For mathematical discussions of oscillations in heat kernel estimates see \cite{Hambly2010,Kajino2010}. In particular, there is a  class of fractals where oscillations are related to large gaps in the spectrum. This topic is a subject of active current research \cite[and references therein]{St2005gaps,HSTZ}. 

\subsection{Spectral volume - Anomalous walk dimension}

We have seen that for a $d$-dimensions Euclidean manifold, the dominant contribution to the Weyl expansion takes the form (\ref{weyld}), {\it i.e.}, it is proportional to the volume $V$ of the manifold. This dependence, as has been emphasized, is a consequence of the fact that the Laplacian sets units of length. In other words, the quantity $t^{d/2} \mbox{Tr} \left( e^{- t \, \Delta} \right)$ has a well defined limit, $V$, for $t \rightarrow 0$. This limit can be viewed as $V = L^d$ where $L$ is the characteristic length set by the Laplacian. There is another consequence of the form $L^d / t^{d/2}$ of the dominant term of the Weyl expansion (\ref{weyld}). Since $Z(t)$ is dimensionless, this imposes the usual Euclidean dispersion $L^2 \propto t$ characteristic of a diffusion processes where units are matched by means of the diffusion coefficient $D$. This can be checked directly from the differential equation (\ref{heat1}) or from its solution (\ref{greenheat}). But for diffusion on a fractal there is no similar local equation nor closed expression of the local propagator $P(r,r',t)$. It is nevertheless possible to express an equivalent dispersion which is solely a property of the Laplace  operator. 

To that purpose, we note from (\ref{zdiam2}), that the quantity $t^{d_s / 2} \,   Z_\Diamond / G_\Diamond $ has a finite limit for $t \rightarrow 0$ which we denoted $V_s$. Restoring length and time units (through $D$) and comparing to (\ref{heat17}) for the Euclidean case, leads necessarily to the form $V_s = L_s ^{d_h}$ where $L_s$ is the characteristic spectral length  associated to the Laplacian on the diamond fractal and $d_h$ is the Hausdorff geometric dimension. Inserting this expression into $Z_\Diamond$ allows to rewrite the power law leading prefactor under the form $ \left( L_s ^{2 \, d_h / d_s} / t \right)^{d_s / 2}$. Since it is a dimensionless quantity, this implies that the dispersion for fractals is of the form $L^2 \propto t^{d_s / d_h}$. This new dispersion plays an essential role in the characterization of diffusion on fractals. It is a constitutive equation usually written under the form of a characteristic mean square displacement law,
\begin{equation}
\langle r^2 (t) \rangle \propto t^{2 / d_w} 
\label{dw1}
\end{equation}
where the exponent $d_w \equiv 2 d_h / d_s$ is called "anomalous walk dimension". 
Thus, the expression,
\begin{equation}
Z_\Diamond (t) = {L_s ^{d_h} \over t^{d_s / 2} } \,  G_\Diamond \left( { 2 \pi \, \ln t \over d_w \, \ln l } \right) = \left( {L_s ^{d_w} \over t} \right)^{d_s / 2} \, G_\Diamond \left( { 2 \pi \, \ln t \over d_w \, \ln l } \right) 
\label{zvs}
\end{equation}
for the heat kernel on a fractal takes a form analogous to its Euclidean counterpart (\ref{heat1}) but where the geometric volume $V = L^d$ is now replaced by a {\it spectral volume} $V_s = L_s ^{d_h}$ which results directly from the spectral properties of the Laplacian. It is to be noted that the {\it geometric volume} of a fractal becomes infinite so that it is unlikely to appear in the Weyl expansion of $t^{d_s/2} \, Z_\Diamond / G_\Diamond$ which has a finite limit for $t \rightarrow 0$.

\section{Thermodynamics on photons : The fractal blackbody \cite{eaprl}}

\subsection{Thermodynamics of the quantum radiation - Generalities}

 A direct application of the considerations developed in the previous sections is to the study of statistical mechanics of  quantum radiation (photons) in fractal structures. Indeed, a basic aspect of it, namely the blackbody radiation precisely addresses the relation between the electromagnetic modes inside a cavity (a manifold) and its geometric characteristics. Historically, the seminal work of H. Weyl which led, among other results, to the expansion  (\ref{weyld}) has been motivated by considerations raised by H. Lorentz about the dependence of the Jean's radiation law upon the volume of the cavity at the exclusion of other geometric characteristics. 
 
 Although the derivation of the thermodynamic partition function of quantum radiation constitutes well known textbook materials, we wish to re-examine it towards its application to fractals in order to emphasize some key points and basic assumptions. The purpose of this derivation is to show that the partition function is directly related to the spectral zeta function so that it can be calculated for Euclidean and fractal manifolds as well. Moreover the geometric information about a manifold retrieved from thermodynamics relates directly to the spectral geometry of the Laplace operator. This point is not always emphasized in textbooks which prefer to consider combinatorics of mode counting in simple cubic geometries, an approach which relies heavily on the existence of Fourier transform and phase space quantization cells, a tool which is not available for fractals. 
 
For convenience, we consider thermodynamics of a scalar massless field (scalar photons) which simplifies the calculation due to the vanishing chemical potential resulting  from zero mass. Generalization to massive fields is possible though slightly more cumbersome \cite{chen,eador}. 
The spectral partition function (including the zero mode Casimir contribution) of a quantum oscillator of frequency $\omega$ at temperature $T = 1 / k_B \beta$ is
$\ln Z (T,\omega)  = - { \beta \hbar \omega \over 2} - \ln \left( 1 - e^{- \beta \hbar \omega} \right)$. Using the identity
\begin{equation}
- \ln \left( 1 - e^{- \beta \hbar \omega} \right) = {\hbar \beta \over 4 \pi} \,  \int_0^\infty {d \tau \over \tau^{3/2}} \, e^{- \omega^2 \tau} \, \sum_{n=1}^\infty e^{- n^2 \, \left( \hbar \beta \right)^2 / 4 \tau}
\end{equation}
together with the Poisson formula
\begin{equation}
\sum_{n= -\infty}^{+ \infty} e^{- n^2 \, \tau} = \sqrt{ { \pi \over \tau}} \sum_{n= -\infty}^{+ \infty} e^{- \pi^2 \, n^2 / \tau}
\label{poisson}
\end{equation}
leads to 
\begin{equation}
\ln Z (T,\omega)  = {1 \over 2} \, \int_0^\infty {d \tau \over \tau} \, e^{- \omega^2 \, \tau}  \sum_{n= -\infty}^{+ \infty} e^{-  \left( {2 \pi n \over \hbar \beta} \right)^2 \, \tau} \, .
\end{equation}
The operator $\partial_0 ^2 \equiv \partial_t ^2$ defined with periodic boundary conditions $\phi (t + \hbar \beta) = \phi (t)$ admits a discrete spectrum $M$ known as Matsubara frequencies $2 \pi n / \hbar \beta$, so that
\begin{equation}
\sum_{n= -\infty}^{+ \infty} e^{-  \left( {2 \pi n \over \hbar \beta} \right)^2 \, \tau} = \mbox{Tr}_M \left( e^{- \tau \partial_0 ^2} \right) \, .
\end{equation}
Identifiying $\omega^2=c^2 k^2$ with the eigenvalues of $c^2 \Delta$, and tracing over all modes,  
we obtain for the partition function of the quantum radiation at temperature $T = 1/ k_B \beta$, 
\begin{equation}
\ln \mathcal Z (T,V)  = {1 \over 2} \, \int_0^\infty {d \tau \over \tau}  \mbox{Tr}_{\mathcal M}  \left( e^{- \tau c^2 \Delta} \right)\,  \mbox{Tr}_{M} \left( e^{- \tau \partial_0 ^2} \right) 
\label{eq6}
\end{equation}
where $\mbox{Tr}_{\mathcal M}$ denotes the trace over the modes of the Laplacian defined on the manifold $\mathcal M$ which contains the anticipated dependence on the volume $V$. 
Finally using the identity $\ln \mathcal O=-\int_0^\infty \frac{d\tau}{\tau}e^{-\mathcal O\, \tau}$, and $\ln {\rm Det}\, \mathcal O={\rm Tr}\,\ln  \mathcal O$, we obtain the elegant and compact dimensionless form for the partition function,
\begin{eqnarray}
 \ln \mathcal Z (T,V) &=& -\frac{1}{2} \ln \mbox{Det}_{M \times \mathcal M} \left( \tilde \partial_0 ^2 + L_\beta^2 \, \Delta \right) \\ 
&=& {1 \over 2} \, \int_0^\infty {d \tau \over \tau}  f (\tau) \mbox{Tr}_{\mathcal M}  \left( e^{- \tau L_\beta ^2 \Delta} \right) 
\label{part2}
\end{eqnarray}
where we have defined the photon thermal wavelength $L_\beta \equiv \hbar  \beta  c$ and the dimensionless operator $\tilde \partial_0 ^2 = \left( \hbar \beta \right)^2 \partial_0 ^2$, so that $f(\tau) =  \mbox{Tr}_{M} \left( e^{- \tau \tilde \partial_0 ^2} \right) = \sum_{n = - \infty}^{+ \infty} e^{- (2 \pi n)^2 \tau}$ is a dimensionless Jacobi $\theta_3$-function. 

From this last expression of the partition function, we see that it is indeed determined by the heat kernel of the Laplace operator on the manifold $\mathcal M$ 
\begin{equation}
\mathcal K_{\mathcal M} (\tau) \equiv {\rm Tr}_{\mathcal M} e^{- \tau \, L_\beta^2 \, {\Delta}} \, .
\label{hk}
\end{equation}
This form is useful to describe in a closed way the thermodynamics of quantum radiation in a large hypercube of volume $L^d$, in $d$ space dimensions. There the modes $\vec k$ are quantized in units of $ 2 \pi {\vec n} / V^{1/d}$ where $\vec n$ is a  vector with integers components defining elementary cells in the reciprocal phase space. The previous expression of the partition function becomes 
\begin{equation}
  \ln \mathcal Z (\beta , V= L^d) = \ln \mathcal Z (\hbar \beta c V^{-1/d}) = -\frac{1}{2} \ln \mbox{Det}_{M \times \mathcal M} \left( \tilde \partial_0 ^2 +  L_\beta ^2 V^{-2/d}  \tilde \Delta \right) \, ,
  \label{cube}
  \end{equation} where $- \tilde \Delta$ is a dimensionless Laplacian. Therefore $\mathcal Z$ is a function of the single variable $L_\beta V^{-1/d}$.  Standard thermodynamic quantities follow immediately from this specific scaling behavior. For example, the equation of state, $P V = U/d$, relating the  internal energy $U$ of the radiation to its pressure $P$ and the volume $V$, follows immediately from
\begin{equation}
U = - {\partial \over \partial \beta} \ln \mathcal Z (T,V) = - \left( {d \ln \mathcal Z (x) \over d x} \right)  \hbar \, c \,  V^{- 1/d}
\label{eqU}
\end{equation}
and 
\begin{equation}
P = {1 \over \beta} \left( {\partial \ln \mathcal Z \over \partial V} \right)_T = - \left( {d \ln \mathcal Z (x) \over d x} \right) {\hbar c  V^{- 1/d} \over V \, d} \, \ .
\label{pressure}
\end{equation}
The Stefan-Boltzmann law for the internal energy $U$ is a consequence of the equation of state and the thermodynamic relation $
 \left( {\partial U \over \partial V} \right)_T = T  \left( {\partial P \over \partial T} \right)_V - P$, while noticing from (\ref{pressure}) that $P$ depends on $T$ only, in the thermodynamic limit. We then obtain $U = a V T^{d+1}$ where $a$ is a constant to be determined from (\ref{cube}). It is already apparent from this simple case that the volume dependence in the thermodynamic equation of state, $P V = U/d$, comes from spectrum of the Laplacian. 
 
  For black-body radiation associated to Euclidean manifolds of complicated shape, it is difficult to make an explicit mode decomposition and  find an explicit expression like (\ref{cube}) for the heat kernel. However, we can learn about the thermodynamic [large volume] limit from the Weyl expansion (\ref{weyld}) of the heat kernel. Note that the large volume limit corresponds to $V \gg  L_\beta ^d$, which is a "high temperature" limit $ k_B T \gg \hbar c /V^{1/d}$. Keeping only the dominant  volume term in (\ref{weyld}), expression (\ref{part2})  leads immediately to the familiar thermodynamic expressions \cite{bookmecastat} previously derived: $\ln\mathcal Z=(V/L_\beta^d) \zeta_R(d+1)\Gamma\left(\frac{d+1}{2}\right)/\pi^{(d+1)/2}$, $P=(k_B T /L_\beta^d) \zeta_R(d+1)\Gamma\left(\frac{d+1}{2}\right)/\pi^{(d+1)/2}$. 
Away from the thermodynamic  limit, subdominant terms in (\ref{weyld}) lead to corrections that depend on the exact geometry of the volume enclosing the radiation, but the equation of state $PV=U/d$ is always valid \cite{baltes76}.
This formulation in terms of heat kernel makes it clear that the Weyl expansion is directly related to the thermodynamic limit of a black-body radiation system, so we can use the leading term as a {\it definition} of the volume probed by the photons as they attain thermal equilibrium.

\subsection{Thermodynamics of the quantum radiation on fractals}

Based on previous results and especially expression (\ref{part2}) of the partition function, we are now in position to study  thermodynamics of quantum radiation on fractals \cite{eaprl}. The heat kernel ${\mathcal K}_{\mathcal F} (\tau)$ equivalent of (\ref{hk}) but on a fractal $\mathcal F$ is obtained using (\ref{zvs}), so that 
\begin{equation}
 {\mathcal K}_{\mathcal F} (\tau) = \mbox{Tr}_{\mathcal F} \left( e^{- L_\beta ^2 \tau \, \Delta} \right) = { \left( L_s / L_\beta \right)^{d_h} \over \tau^{d_s / 2}} \, G_{\mathcal F} \left( { 2 \pi \, \ln \tau \over d_w \, \ln l } \right) \, .
 \label{hkthermo}
 \end{equation}
The thermodynamic partition function $\ln \mathcal Z \left( \hbar  \beta c V_s ^{- 1/ d_h} \right)$ is a function of the single variable $L_\beta V_s ^{- 1/ d_h}$, where $V_s ^{ 1/ d_h} = L_s$ is the spectral length associated to the Laplacian on the corresponding fractal $\mathcal F$.  Thermodynamic quantities and relations follow immediately from this specific scaling behavior and from (\ref{eqU}) and (\ref{pressure}) to give the thermodynamic equation of state on a fractal as $P V_s = U / d_h$. 
The second important conclusion is that the actual expressions for pressure $P$, internal energy $U$, etc... will be modified on a fractal, not only by the appearance of the spectral dimension $d_s$ and spectral volume $V_s$, but also by the appearance of oscillatory terms arising from the behavior of the log-periodic function $G_{\mathcal F}$ in (\ref{hkthermo}). 

\subsection{Vacuum Casimir energy}

Another straightforward consequence of (\ref{hkthermo}) is the expression of the zero temperature free energy namely  the vacuum Casimir energy calculated in various complex geometries and recently on quantum graphs \cite{harrison}. A general expression for the Casimir energy is obtained from the inverse Mellin transform of the spectral zeta function which gives
\begin{eqnarray}
 \ln \mathcal Z(T, V)&=&-\frac{1}{2}\left(\frac{L_\beta}{L}\right)\,\zeta_{\mathcal M}\left(-\frac{1}{2}\right)\nonumber\\
&&\hskip -2cm +\frac{1}{\pi i}\int_C \left(\frac{L}{L_\beta}\right)^{2s}\, \Gamma(2s)\, \zeta_{R}(2s+1)\, \zeta_{\mathcal M}(s) \, ds
\label{zetaform}
\end{eqnarray}
The first term gives the standard zero temperature ``vacuum Casimir energy" contribution \cite{dowker}, proportional to $\zeta_{\mathcal M}\left(-\frac{1}{2}\right)$. On a fractal ${\mathcal F}$, we have $L = L_s$ and the previous expression leads for the vacuum Casimir energy to: 
\begin{equation}
E_0= { 1 \over 2} {\hbar c \over L_s} \zeta_{\mathcal F} \left( -\frac {1}{ 2} \right) \, .
\label{casimir}
\end{equation}

\section{Conclusion and some open questions}

We have presented general features of deterministic self similar fractals which possess an exact decimation symmetry. To describe quantitatively diffusion processes and wave propagation on these structures, we have defined and develop  spectral tools related to the Laplacian, such as the heat kernel and the spectral zeta function. After reviewing some of their main features for Euclidean manifolds, we have calculated them on fractals. This has enabled us to single out a number of specific features of fractals such as the spectral dimension $d_s$, the spectral volume $V_s$, the existence of log-periodic oscillations of spectral quantities, an unusual dispersion characterized by the walk dimension $d_w = 2 d_h / d_s$, $d_h$ being the geometric Hausdorff dimension. These features show up when extending results of statistical mechanics and quantum field theory on fractal structures. To see this at work, we have considered the problem of thermodynamics of quantum radiation on fractals in the simplified scalar version as well as the vacuum Casimir energy. The results we have obtained generalize straightforwardly to related problems such as massive bosons.  In that case, it is  a direct consequence of the form of the heat kernel that Bose-Einstein condensation occurs for $d_s > 2$ {\it i.e.} independently of the geometric dimension $d_h$ \cite{chen,eador}, a fact already recognized long ago \cite{rammaljphys}. More involved is the problem of the occurrence of superfluidity (a phenomenon distinct from Bose-Einstein condensation) which depends on terms of higher order than the volume contribution in the Weyl expansion \cite{eador}. Beyond equilibrium situations, the problem of quantum emission of radiation (either spontaneous or stimulated) is highly interesting either from a fundamental point of view or for applications. It can be shown \cite{eagurevich} that the probability of emission, also known as vacuum persistence in the context of quantum field theory, is driven by another spectral kernel, the sinc kernel (\ref{convolution2}) related to the heat kernel and which exhibits an analogous qualitative behavior. 

We have mostly discussed spectral quantities and not the behavior of non-diagonal terms in the propagator $P(x,y,t)$ associated to diffusion. Very little is known about it except for useful bounds \cite{barlowperkins} and a conjectured form supported by numerical results \cite{eavoit}. This part of the characterization of fractals is nonetheless essential to better understand stochastic processes, out of equilibrium statistical mechanics and large deviation physics. Preliminary results based on specific distributions of traps \cite{bernasconi} or on an additivity principle \cite{derrida} seem promising. 

Let us mention finally that the absence of translational invariance in fractals prevents using Fourier transforms. It is preferable instead to use Mellin transform as explained in section 2. This has important consequences in formulating an uncertainty principle relating direct to reciprocal spaces. We have emphasized that while the direct space is driven by the Hausdorff dimension $d_h$, candidates for reciprocal spaces obtained from the Laplacian are instead driven by $d_s$. This issue has far reaching consequences if one wishes to properly formulate canonical field quantization on fractal structures including spin. This leads to yet another challenge on fractals. A proper understanding of spin on Euclidean manifolds has required tools such as heat kernel and Weyl expansion of Dirac operators. Topological properties related to spin thus appear under the form of powerful results known as Index theorems \cite{atiyah}. Their extension to fractals requires a proper understanding of cohomology, connexions and definition of a Laplacian using Hodge theorem \cite{eakirone}.

\subsection*{Acknowledgments} : It is a pleasure to thank G. Dunne,  A. Teplyaev, E. Gurevich, D. Gittelman and O. Spielberg for collaboration and discussions.

\bibliographystyle{amsalpha}

\end{document}